\documentclass[epsf,floatfix,prl,aps,twocolumn]{revtex4}

\usepackage{epsf}
\pdfoutput=1

\usepackage{amsmath,graphicx}

\def\normv{ {\bf e}_{\eta}}
\def\longv{ {\bf e}_{\sigma}}

\begin{document}

\date{\today}

\title{Unified continuum approach to crystal surface morphological relaxation}

\author{Dionisios Margetis}

\affiliation{Department of Mathematics and Institute for Physical Science and Technology, 
University of Maryland, College Park, MD 20742, USA}

\begin{abstract}
A continuum theory is used to predict scaling laws for the morphological relaxation of crystal surfaces 
in two independent space dimensions. The goal is to unify previously disconnected experimental observations
of decaying surface profiles. The continuum description is derived
from the motion of interacting atomic steps. For isotropic diffusion of adatoms across each terrace,
induced adatom fluxes transverse and parallel to step edges obey different laws,
yielding a {\sl tensor} mobility for the continuum surface flux. The partial differential equation (PDE) for the height profile 
expresses an interplay of step energetics and kinetics, and aspect ratio of surface topography that
plausibly unifies observations of decaying bidirectional surface corrugations.
The PDE reduces to known evolution equations for axisymmetric mounds and one-dimensional periodic corrugations.\looseness=-1
\end{abstract}

\maketitle

\section*{}

Novel small devices rely on the stability of 
nanoscale surface features. The lifetimes of nanostructures decaying via surface diffusion 
scale as a large power of their size and increase with decreasing temperature. 
Below roughening, crystal surfaces evolve via the motion of atomic steps 
bounding nanoscale terraces~\cite{bcf51,jeongwilliams99}. 

Experiments with decaying surface 
features~\cite{ichimiyaetal00,thurmeretal01,keefeetal94,blakelyetal97,erlebacheretal00,pedemonteetal03,zhouetal07} 
are useful for testing step models. Particularly informative are observations of bidirectional  
corrugations relaxing below roughening~\cite{keefeetal94,blakelyetal97,erlebacheretal00,pedemonteetal03,zhouetal07}. 
In lithography-based experiments~\cite{keefeetal94}, 
where initial wavelengths in two directions differ significantly and profiles 
depend nearly on one space dimension (1D), the surface height decays exponentially with time. By contrast, 
in sputter-rippling experiments~\cite{erlebacheretal00,pedemonteetal03}, where initial wavelength ratios are closer to unity 
and profiles evidently depend on two space dimensions (2D), height spatial-frequency components 
decay inverse linearly with time. These observations have previously evaded a unified
theory~\cite{israelikandel-prl-comm,erlebacheretal-prl-comm}. 
In this Letter, we use a continuum theory to plausibly unify these
observations via an appropriate {\it tensor mobility}. \looseness=-1

There are two main theoretical approaches to crystal surface morphological evolution below roughening.
One approach follows the motion of steps in the spirit of the Burton-Cabrera-Frank (BCF) model~\cite{bcf51} 
via numerical solutions of coupled equations for step 
positions~\cite{israelikandel00,israelikandel}. Step simulations in 1D~\cite{israelikandel00} show exponential decay 
of surface corrugations with attachment-detachment limited (ADL) kinetics, 
in agreement with lithography experiments~\cite{keefeetal94}. 
Step simulations in 2D invoke axisymmetry~\cite{israelikandel}, and
are thus limited in their ability to make predictions for general surface morphologies.

Another approach relies on equilibrium thermodynamics and mass conservation using continuum evolution 
laws~\cite{rettorivillain88,ozdemirzangwill90,spohn93,margetisetal05,
shenoyfreund02,shenoyetal04,chanetal04} such as partial differential equations (PDEs).
PDEs enable simple scaling predictions; see e.g.~\cite{margetisetal05}. 
Continuum models are criticized~\cite{israelikandel00} for their inaccurate description of macroscopic, 
planar surface regions (``facets''), but progress is made in including facets in 
evolution laws~\cite{margetisetal06}. Continuum theories have not previously unified 
observations of decaying surface corrugations~\cite{erlebacheretal-prl-comm}. An ingredient of such theories
is the {\it scalar} mobility for the adatom flux in 
2D~\cite{israelikandel-prl-comm,margetisetal05,shenoyetal04,chanetal04}, which does not
essentially distinguish adatom fluxes parallel to steps from fluxes transverse to steps.
This formulation is valid when steps are everywhere parallel~\cite{margetisetal05},
but is shown here to be inadequate in general cases.

In this Letter we plausibly unify experimental observations
of decaying profiles by invoking a {\it tensor macroscopic} mobility
for the adatom flux in a setting of {\it isotropic} terrace diffusion;
see Eqs.~(\ref{eq:Mxx})--(\ref{eq:Myy}). 
An elaborate derivation is given elsewhere~\cite{margetiskohn}.
Here, we provide a more general yet simpler derivation. We show that the resulting PDE 
for the height profile reduces to known evolution laws for 1D gratings and 2D nanostructures.
Further, we relate scaling predictions of the general theory to relaxation experiments.
We find that observed scaling laws with time can arise from competition of step kinetics 
with surface topography. This effect is due to coupling of adatom flux components via
terrace diffusion, and is distinct from the influence of step edge diffusion, e.g. work in~\cite{paulinetal01}.
A similar effect of anisotropic terrace diffusion on step meandering is studied in~\cite{dankeretal04}.
By contrast to~\cite{dankeretal04}, our step model has {\it scalar} microscopic parameters. \looseness=-1

First, we describe the model of step flow~\cite{bcf51}. 
A top terrace is surrounded by non-self-intersecting and non-crossing
steps numbered $i=1,\,2,\,\ldots\,$; $i=1$ denotes the top step.
The projection of steps on the basal (high-symmetry) plane is described 
by the position vector ${\bf r}(\eta,\sigma,t)$ where $t$ is time; $\eta=\eta_i$ at the $i$th step,  
$\eta_i < \eta<\eta_{i+1}$ on the $i$th terrace, and $\sigma$ gives the position along each step; see Fig.~1.
The unit vectors normal and parallel to steps
in the direction of increasing $\eta$ and $\sigma$ are $\normv$ and $\longv$;
$\normv\cdot\longv =0$. The metric coefficients (to be used below) are 
$\xi_{\eta}=|\partial_{\eta}{\bf r}|$ and $\xi_{\sigma}=|\partial_{\sigma}{\bf r}|$;
$\partial_\eta:=\partial/\partial\eta$.

\begin{figure}
\includegraphics*[scale=.08,trim=5in 16.6in 1in 5in]{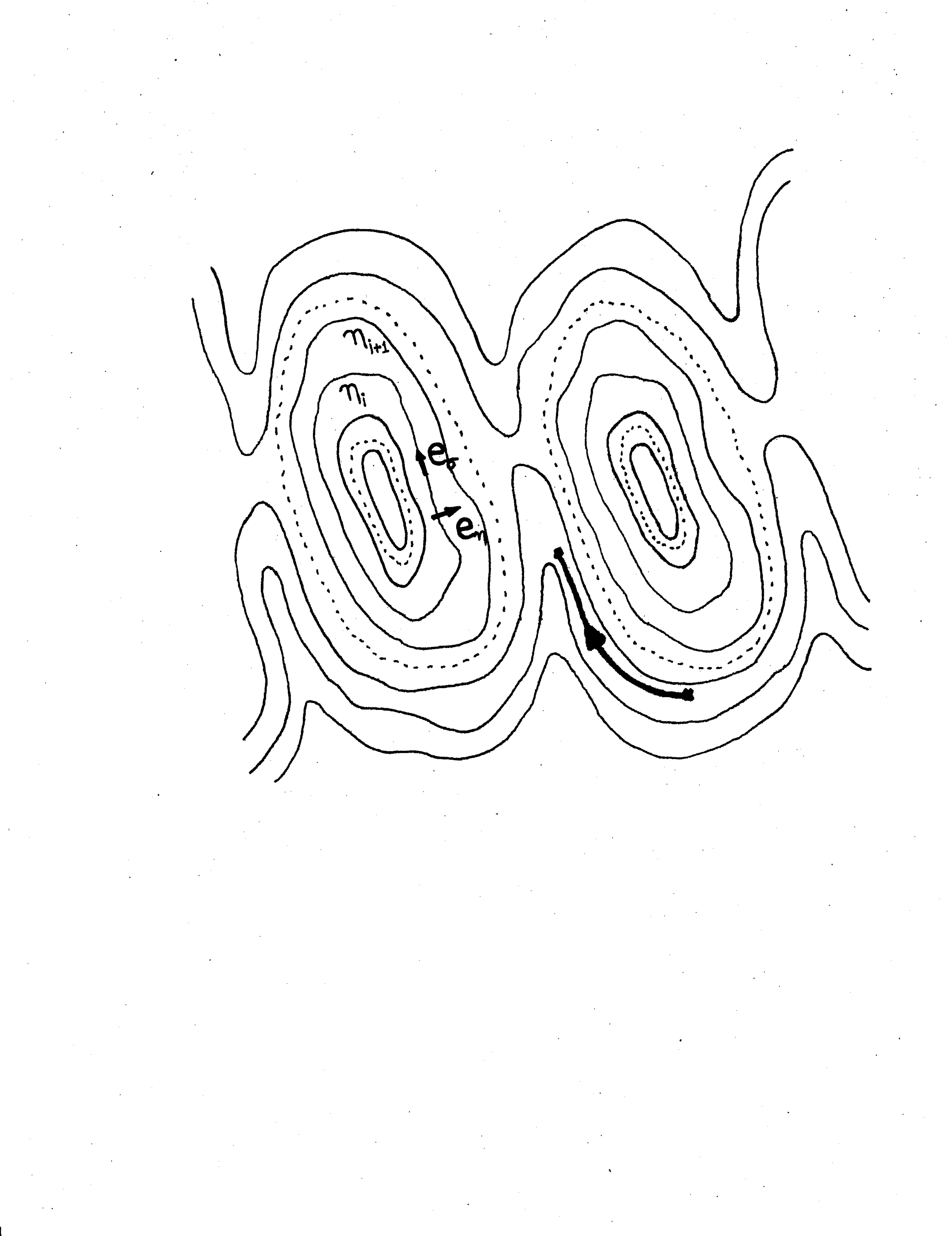}
\caption{Schematic of steps on 
the basal plane. Local coordinates relative to a top terrace are ($\eta,\sigma$).
The arrow shows longitudinal flux directed to a valley. Dots 
denote many steps.\looseness=-1}
\end{figure}

Mass conservation for atoms is described by 
\begin{equation}
v_i=\normv\cdot\ 
\frac{{\rm d}{\bf r}}{{\rm d}t}\bigl|_{\eta=\eta_i}=\frac{\Omega}{a}\,[J_{i-1}^{\eta}(\eta_i,\sigma,t)-J_i^{\eta}(\eta_i,\sigma,t)];
\label{eq:mass-conserv}
\end{equation}$v_i$ is the (normal) velocity of the $i$th step,  
$\Omega$ is the atomic volume, and $a$ is the step height; $J_i^{\eta}={\bf J}_i\cdot\normv$ is 
the adatom current (adatoms/length/time) transverse to steps;
${\bf J}_i=-D_{\rm s}\nabla C_i$ is the adatom current on the $i$th terrace, 
$D_{\rm s}$ is the terrace diffusivity, a scalar function of ${\bf r}$, and
$C_i({\bf r},t)$ is the adatom density [adatoms/(length)$^2$] on the $i$th terrace. 
The variable $C_i$ solves the diffusion equation,
which in the quasistatic approximation becomes $\nabla^2 C_i\approx 0$, where no material is deposited from above. 
The requisite boundary conditions describe atom attachment-detachment at the $i$th and $(i+1)$th steps~\cite{israelikandel},
\begin{equation}
\mp J_i^{\eta}(\eta_{l},\sigma')=k \,[C_i(\eta_{l},\sigma')-C_{i}^{\rm eq}(\sigma')].\label{eq:bcs}
\end{equation}
The time ($t$-) dependence is omitted, $l=i$ (upper sign) or $i+1$ (lower sign), 
$k$ is the attachment-detachment rate, and $C_i^{\rm eq}$ is the $i$th step equilibrium atom density.
Note that Eqs.~(\ref{eq:bcs}) are similar to those appearing in other
growth problems; but in the present case there is no morphological instability.\looseness=-1 

Next, we close Eqs.\ (\ref{eq:mass-conserv}) and (\ref{eq:bcs}) by relating $C_i^{\rm eq}$ with the step positions. First, we introduce the step chemical potential of the $i$th step, 
$\mu_i(\sigma,t)$, the change in the step energy by adding or removing an atom at $(\eta_i,\sigma)$~\cite{israelikandel}:  
$C_i^{\rm eq}=C_{\rm s}\,e^{\mu_i/(k_{\rm B}T)}\sim C_{\rm s}\bigl( 1+\frac{\mu_i}{k_{\rm B}T}\bigr)$, where
$|\mu_i|\ll k_{\rm B}T$,
$C_{\rm s}$ is the atom equilibrium density near a straight isolated step and 
$k_{\rm B}T$ is the Boltzmann energy. 

Second, we provide a relation of $\mu_i$ with the step positions. We use $U(\eta,\sigma)$, the energy of atoms 
per length of the $i$th step (for $\eta=\eta_i$); 
thus, the length $\delta s_i=\xi_{\sigma}\delta\sigma$ of the $i$th step has energy
$\delta W_i=U(\eta_i,\sigma)\delta s_i$. Addition or removal of atoms at $(\eta_i,\sigma)$ causes 
$\eta_i$ to change by $\delta\eta$ assuming energy isotropy, the step to move along the local normal ($\normv$) by distance
$\delta\varrho=\xi_{\eta}\delta\eta$, and the step energy to change by 
$\delta^2 W_i=[\partial_{\eta_i}(\delta W_i)]\delta\eta$. By definition of $\mu_i$, 
$\mu_i=\frac{\Omega}{a}\frac{\delta^2 W_i}{\delta\varrho\,\delta s_i}$ for $(\delta\sigma,\delta\eta)\to 0$, we find 
\begin{equation}
\mu_i=(\Omega/a)[(\partial_{\eta}U)\,\xi_{\eta}^{-1}+\kappa\,U(\eta,\sigma)]|_{\eta=\eta_i},
\label{eq:mu-def}
\end{equation}where $\kappa$ is the step curvature and
$U=\gamma+U^{\rm int}$; $\gamma$ is the step line tension, assumed a constant, 
and $U^{\rm int}$ accounts for interactions with other steps. 
For nearest-neighbor elastic-dipole or entropic repulsions, $U^{\rm int}$ is~\cite{jeongwilliams99,marchenkoparshin}
\begin{equation}
U^{\rm int}=g\,\biggl[\frac{\Phi(\eta,\eta_{i+1};\sigma)}{(\eta_{i+1}-\eta)^{2}}
+\frac{\Phi(\eta,\eta_{i-1};\sigma)}{(\eta-\eta_{i-1})^{2}}\biggr],\label{eq:local-inter}
\end{equation}where $g$ (energy/length) is positive, and $\Phi(\eta,\zeta;\sigma)$ is geometry-dependent,
differentiable with $(\eta,\zeta)$ and satisfies
$\Phi(\eta_i,\eta_{i+1})\delta s_i=\Phi(\eta_{i+1},\eta_i)\delta s_{i+1}$~\cite{margetiskohn}. 
Suppressing $\eta_{i-1}$ and $\eta_{i+1}$, Eqs.\ (\ref{eq:mu-def}) and (\ref{eq:local-inter}) yield 
$\mu_i= \tilde\mu(\eta=\eta_i,\sigma,t)$.
 
Equations (\ref{eq:mass-conserv})--(\ref{eq:local-inter}) describe coupled step motion via adatom isotropic diffusion across
terraces and atom attachment-detachment at steps. To enable predictions 
for decaying surface profiles at length scales large compared to the terrace width, $\delta\varrho_i$, 
we next derive a PDE for the continuum height profile, $h({\bf r},t)$. 
Thus, $\delta\varrho_i$ is small compared to: (i) the length  
over which the step density, $\frac{a}{\delta\varrho_i}$, varies; and (ii) the step radius of curvature, $1/\kappa$. 
We take $\delta\eta_i=\eta_{i+1}-\eta_i\to 0$ with fixed $\frac{a}{\delta\varrho_i}$. In this limit, 
$\frac{a}{\delta\rho_i}\to |\nabla h|$, where $\nabla h=(\partial_x h,\partial_y h)$, and 
$v_i\to\frac{\partial_t h}{|\nabla h|}$.

First, we note that the familiar continuum mass conservation statement for atoms comes from the step velocity law, 
Eq.\ (\ref{eq:mass-conserv}). By using the continuum surface current ${\bf J}({\bf r},t)$, the continuous extension of 
${\bf J}_i(\eta_i,\sigma,t)$, we obtain
\begin{equation}
\partial_t h= -\frac{\Omega}{\xi_{\sigma}\,\xi_{\eta}}[ \partial_{\eta}(\xi_{\sigma} J^{\eta})
+\partial_{\sigma}(\xi_{\eta} J^{\sigma})]=-\Omega\nabla\cdot {\bf J}.\label{eq:conserv-full}
\end{equation}

Next, we apply Eqs.\ (\ref{eq:bcs}) to relate ${\bf J}({\bf r},t)$ to the continuum step chemical
potential, $\mu({\bf r},t)=\tilde\mu(\eta_i,\sigma,t)$. The following procedure is more
general than the analysis in~\cite{margetiskohn}.
(i) We apply Eq.\ (\ref{eq:bcs}) with the upper sign for $\sigma'=\sigma$, 
and with the lower sign for $\sigma'=\sigma+\delta\sigma$.  
(ii) We expand the transverse current, $J^{\eta}_i$, the density $C_i$ and $\tilde\mu$, each
evaluated at $(\eta_{i+1},\sigma+\delta\sigma$), at ($\eta_i,\sigma$) using ${\bf J}_i=-D_{\rm s}\nabla C_i$,
e.g., $C_i|_{i+1,\sigma+\delta\sigma}\approx C_i|_{i}-D_{\rm s}^{-1}(\xi_\eta\delta\eta_i
J_i^{\eta}|_{i}+\xi_\sigma\delta\sigma J_i^{\sigma}|_{i})$ where $Q|_p:=Q(\eta_p,\sigma)$
and $J_i^\sigma={\bf J}_i\cdot\longv$ is the longitudinal current. (iii) We subtract Eqs.\ (\ref{eq:bcs}) 
dropping terms that are negligible as $\delta\eta_i\to 0$. Thus, we find
\begin{equation}
\biggl( 1+q\frac{a}{\delta\rho_i}\biggr) J_{i}^{\eta}
+\frac{C_{\rm s}D_{\rm s}}{k_{\rm B}T}\frac{\partial_{\eta}\tilde\mu}{\xi_{\eta}}
+\frac{\xi_{\sigma}}{\xi_{\eta}}\biggl(J_i^{\sigma}+\frac{C_{\rm s}D_{\rm s}}{k_{\rm B}T}
\frac{\partial_{\sigma}\tilde\mu}{\xi_{\sigma}}\biggr)\delta\sigma=0,
\label{eq:bcs-expandII}
\end{equation}
where $q=\frac{2D_{\rm s}}{ka}$. By setting $\delta\sigma=0$ in Eq.\ (\ref{eq:bcs-expandII}) we obtain
\begin{subequations}
\label{alleqs:J-mu}
\begin{equation}
J_i^{\eta}\rightarrow {\bf J}({\bf r},t)\cdot \normv = -\frac{D_{\rm s} C_{\rm s}}{k_{\rm B}T}\,
\frac{1}{{\displaystyle
1+q\,|\nabla h|}}\frac{\partial_\eta \mu}{\xi_\eta},\label{eq:J-mu-normal}
\end{equation}where $q(a/\delta\rho_i)$ is fixed. Hence, Eq.\ (\ref{eq:bcs-expandII}) reduces to
\begin{equation}
J_i^{\sigma}\rightarrow 
{\bf J}({\bf r},t)\cdot \longv =-\frac{D_{\rm s} C_{\rm s}}{k_{\rm B}T}\frac{\partial_\sigma \mu}{\xi_\sigma}.
\label{eq:J-mu-longit}
\end{equation}
\end{subequations}

By Eq.\ (\ref{eq:J-mu-longit}), the continuum longitudinal current, $J^{\sigma}$, 
has the terrace diffusivity, $D_{\rm s}$, whereas the normal current $J^{\eta}$, 
Eq.\ (\ref{eq:J-mu-normal}), has the slope-dependent effective diffusivity 
$\tilde D_{\rm s}=D_{\rm s} (1+q|\nabla h|)^{-1}$; $\tilde D_{\rm s}$ 
equals $D_{\rm s}$ for terrace-diffusion limited (TDL) kinetics, $q |\nabla h|\ll 1$. 
This behavior results from coarse-graining in 2D, 
combining atom attachment-detachment, terrace diffusion and 
step topography. For ADL kinetics, $q|\nabla h| \gg 1$, $J^{\eta}$ is sensitive to {\it step} variations of $\mu$ 
because steps are sources and sinks of atoms by Eqs.\ (\ref{eq:bcs}), whereas $J^{\sigma}$ is sensitive to 
{\it space} variations of $\mu$ along steps due to adatom diffusion between non-parallel steps. 
Equations (\ref{alleqs:J-mu}) read ${\bf J}=-C_{\rm s}{\bf M}\cdot \nabla\mu$ where the mobility ${\bf M}$ 
(length$^2$/energy/time) is a second-rank {\it tensor} (a $2\times 2$ matrix where
${\bf J}$ and $\nabla\mu$ are 2-column vectors). In the basal's plane Cartesian system ($x,\,y$)
the matrix elements $M_{ij}$ ($i,\,j= x,\,y$) of ${\bf M}$ are  
\begin{eqnarray}
M_{xx}&=&\frac{D_{\rm s}}{k_{\rm B}T}\frac{(\partial_x h)^2}{|\nabla h|^2}\biggl[
\frac{1}{1+q|\nabla h|}+\alpha^2\biggr],\label{eq:Mxx}\\ 
M_{xy}&=&M_{yx}= -\frac{D_{\rm s}}{k_{\rm B}T}\frac{q|\nabla h|}{1+q|\nabla h|}
\frac{(\partial_x h)^2}{|\nabla h|^2}\ \alpha,\label{eq:Mxy}\\
M_{yy}&=&\frac{D_{\rm s}}{k_{\rm B}T}\frac{(\partial_x h)^2}{|\nabla h|^2}
\biggl[\frac{\alpha^2}{1+q|\nabla h|}+1\biggr],\label{eq:Myy}
\end{eqnarray}where $\alpha:=\frac{\partial_y h}{\partial_x h}$. For biperiodic
profiles, $\alpha$ is estimated by $\frac{\lambda_x}{\lambda_y}$, the (aspect) ratio of dominant (maximum-amplitude) 
wavelengths in $x$ and $y$; we take $\lambda_x \le\lambda_y$ and, thus, $\alpha \le 1$.

Next, we obtain a PDE for the height profile, $h({\bf r},t)$. First, we derive a relation of $\mu$ with $\nabla h$ via 
Eqs.\ (\ref{eq:mu-def}) and (\ref{eq:local-inter}).
(i) We expand in $(\eta_i-\zeta)$ the function $\Phi(\eta_i,\zeta;\sigma)$ of Eq.\ (\ref{eq:local-inter}), 
where $\zeta=\eta_{i+1}$ or $\eta_{i-1}$. (ii) We use an identity for $\Phi$, which stems from the
definition of $U^{\rm int}$~\cite{margetiskohn}. After some algebra, the limit
$\delta\eta_i\to 0$ yields 
\begin{equation}
\mu = \Omega[g_1\,\kappa - g_3\,\nabla\cdot(|\nabla h|\nabla h)],
\label{eq:mu-continuum2}
\end{equation}where $\kappa=-\nabla\cdot\frac{\nabla h}{|\nabla h|}$ is the step edge curvature, 
$g_1=\frac{\gamma}{a}$ and $g_3=\frac{3g}{a}(\frac{\xi_{\eta}}{a})^2 \Phi(\eta_i,\eta_i)$; 
$g_1$ and $g_3$ have dimensions energy per area. This $\mu$ also results
from the variational derivative of the surface energy 
$E=\int\!\int {\rm d}x{\rm d}y\,[g_1|\nabla h|+(g_3/3)|\nabla h|^3]$~\cite{margetisetal05,margetiskohn}.
By Eqs.\ (\ref{eq:conserv-full}), (\ref{alleqs:J-mu}) and 
(\ref{eq:mu-continuum2}), 
\begin{equation}
\partial_t h= B\nabla\cdot\Biggl\{{\bf \Lambda}\cdot\nabla\Biggl
[\nabla\cdot\Biggl(\frac{\nabla h}{|\nabla h|}\Biggr)+ \frac{g_3}{g_1}
\nabla\cdot(|\nabla h|\nabla h)\Biggr]\Biggr\},\label{eq:full-PDE}
\end{equation}where ${\bf \Lambda}=-\frac{k_{\rm B}T}{D_{\rm s}}{\bf M}$ 
and $B=\frac{D_{\rm s}C_{\rm s}g_1\Omega^2}{k_{\rm B}T}$ [(length)$^4$/time].
By Eqs.\ (\ref{eq:Mxx})--(\ref{eq:Myy}) for ${\bf M}$, Eq.\ (\ref{eq:full-PDE}) 
describes an interplay of step energetics and kinetics, 
and aspect ratio $\alpha$. This dependence on $\alpha$ is absent in previous 
studies of morphological evolution below roughening~\cite{israelikandel-prl-comm,margetisetal05,shenoyetal04,chanetal04}.\looseness=-1

It is tempting to compare Eq.~(\ref{eq:full-PDE}) and its ingredients
to similar continuum laws for steps,
e.g. Eq.~(14) of~\cite{dankeretal04} for a step meander without deposition.
The last term of Eq.~(14) in~\cite{dankeretal04} pertains to the flux along the step edge, with
a mobility that depends on the step edge slope. In the small slope limit, this term appears
to agree with Eq.~(\ref{eq:J-mu-longit}). We emphasize that the isotropic physics of our model is different
from that of~\cite{dankeretal04} where anisotropic terrace diffusion coexists with
step edge diffusion.  

We now show that Eq.~(\ref{eq:full-PDE}) reduces properly to known macroscopic laws for everywhere parallel
steps. First, we have $J^\sigma=0$ by which the effective mobility becomes 
$M=\frac{D_{\rm s}}{k_{\rm B}T}(1+q|\nabla h|)^{-1}$, a scalar. For straight steps (in 1D), $\eta=x$, we have $\kappa\equiv 0$ and the
PDE becomes $\partial_t h=-B_3\partial_x[(1+q |\partial_x h|)^{-1}\partial_{xx}(|\partial_x h|\partial_x h)]$
where $B_3=\frac{D_{\rm s}C_{\rm s}g_3\Omega^2}{k_{\rm B}T}$,
which is consistent, for example, with~\cite{israelikandel00}. The reduced PDE can be applied
to systems of periodic corrugations in 1D~\cite{keefeetal94,israelikandel00,shenoyfreund02}.
For concentric circular, descending steps in 2D, $\eta\propto r$ (polar distance), we have $\kappa=1/r$
and the PDE~(\ref{eq:full-PDE}) becomes 
$\partial_t h=B r^{-1}\partial_r\{(1+q m)^{-1}[-r^{-1}+\frac{g_3}{g_1}r\partial_r (r^{-1}\partial_r(rm^2))]\}$
where $m=|\partial_r h|$, which is applied to decaying axisymmetric 
mounds~\cite{thurmeretal01,israelikandel,margetisetal05,margetisetal06}.\looseness=-1

We now apply separation of variables to Eq.~(\ref{eq:full-PDE}) for smooth regions, aiming
to unify decay laws in relaxation experiments. Consistent with step 
simulations in 1D~\cite{israelikandel00} and kinetic Monte Carlo simulations in 
2D~\cite{shenoyetal04}, both for initial sinusoidal profiles, we set
$h({\bf r},t)\approx A(t) H({\bf r})$
and find $A(t)$. This variable separation, which we call a ``scaling Ansatz'',
is satisfied only approximately: additive terms in $\mu$ and ${\bf M}$ scale differently with $A$.  
In $\mu$, Eq.~(\ref{eq:mu-continuum2}), the step line tension ($g_1$ term) scales with $A^0$ and
the step interaction ($g_3$ term) scales with $A^2$; in $M_{ij}$, 
Eqs.\ (\ref{eq:Mxx})--(\ref{eq:Myy}), the kinetic term $\beta=(1+q\langle|\nabla h|\rangle)^{-1}$
must be compared to the aspect ratio squared, $\alpha^2$; 
$\langle |\nabla h|\rangle \equiv\nu$ is a typical slope.\looseness=-1

Our analysis does not address the evaluation of $H({\bf r})$, which solves a nonlinear PDE. 
Because boundary conditions for $H$ at facet edges
require feedback from step simulations~\cite{margetisetal06}, a viable numerical
scheme for $H$ is not possible at the moment. 
By~\cite{israelikandel00,shenoyetal04}, the scaling Ansatz seems reasonable for long $t$
and initial sinusoidal profiles.

We next focus on ADL kinetics, $\beta\ll 1$, distinguishing four cases.
In the first case: (i) step interactions dominate,  
$|g_3\nabla\cdot(|\nabla h|\nabla h)| \gg | g_1\kappa |$ or 
$\frac{g_3}{g_1} \gg (\frac{\alpha}{\nu})^2$ by dimensional analysis for sinusoidal profiles, where 
$\nu\approx\frac{h_{\rm pv}}{\lambda /2}$ and $h_{\rm pv}$ is the peak-to-valley height variation; 
and (ii) $\beta< \alpha^2$ so that longitudinal fluxes are considerable.
Thus, $\mu$ scales with $A^2$, and the matrix elements 
of ${\bf \Lambda}$ are $\Lambda_{xx}\approx -\frac{(\partial_y h)^2}{|\nabla h|^2}$, 
$\Lambda_{yy}\approx -\frac{(\partial_x h)^2}{|\nabla h|^2}$,
and $\Lambda_{xy}=\Lambda_{yx}\approx \frac{(\partial_x h)(\partial_y h)}{|\nabla h|^2}$, which scale with $A^0$
as in TDL kinetics. We find ${\dot A}\approx -c B_3\,A^2$, where 
the dot denotes time derivative. Hence,\looseness=-1
\begin{equation}
A(t)=A_0\ (1+cB_3A_0\, t)^{-1};\quad A_0:=A(0).\label{eq:A-I}
\end{equation}
The constant parameter $c$ [(length)$^{-4}$] 
depends on $H({\bf r})$ and is thus affected by facet evolution. 
Equation (\ref{eq:A-I}) suggests that 
surface relaxation is inverse linear with time if the ($y$-) adatom flux in the direction of the longer
wavelength ($\lambda_y$) is significant.\looseness=-1

In the second case: (i) step interactions remain dominant, 
and (ii) $\beta >\alpha^2$, so that transverse fluxes prevail.
Thus, we obtain ${\dot A}=-{\mathcal C}B_3A$, by which
\begin{equation}
A(t)=A_0\ e^{-{\mathcal C}B_3 t},\label{eq:A-III}
\end{equation}where $\mathcal C$ is affected by $H$.
The remaining cases for ADL kinetics follow similarly. The results are
summarized in Table~I. The square-root decay with time
when line tension dominates and $\beta > \alpha^2$ is in agreement with~\cite{shenoyetal04}.\looseness=-1
\begin{table}[htbp]
\caption{Summary of decay laws for the height amplitude $A(t)$ in ADL kinetics, $\beta\ll 1$.
The top row lists the possible kinetic-geometric conditions on the mobility, Eqs.~(\ref{eq:Mxx})--(\ref{eq:Myy}). 
The leftmost column lists the possible dominant effects in $\mu$, Eq.~(\ref{eq:mu-continuum2}).
The parameter $t^*$ ($t^*>t$) depends on $A(0)$ and $H$.} 
\begin{tabular}{p{1.05in}|p{1.10in}p{1.05in}}
\hline\hline
&  $\alpha^2\ll \beta\ll 1$  & $\beta\ll \alpha^2\le 1$ \\
\hline 
Step interaction &  $A_0 e^{-\mathcal C B_3 t}$& $A_0(1+cB_3A_0 t)^{-1}$\\
Line tension & $A_0\sqrt{1-t/t^*}$ & $A_0(1-t/t^*)$\\
\hline\hline
\end{tabular}
\label{table1}
\end{table}

Our predictions, based on Eq.\ (\ref{eq:full-PDE}) with ADL kinetics, 
can be extended to TDL kinetics. The mobility ${\bf M}$  then
reduces to $\frac{D_{\rm s}}{k_{\rm B}T}$. Thus, we obtain
(\ref{eq:A-I}) or (\ref{eq:A-III}), regardless of $\alpha$, for step-interaction or line-tension dominated $\mu$.

Next, we compare our predictions with observations of 
Si(001)~\cite{keefeetal94,erlebacheretal00} and Ag(110)~\cite{pedemonteetal03} corrugations.
In Si(001), with $\ell\equiv 2D_{\rm s}/k \gtrsim 1000$ nm~\cite{israelikandel-prl-comm} and 
terrace width $\Delta w\lesssim 10$ nm~\cite{keefeetal94,erlebacheretal00,israelikandel-prl-comm},
$q\langle|\nabla h|\rangle\approx \frac{\ell}{\Delta w}\gtrsim 100$ which suggests ADL kinetics. 
We find decay laws comparing (i) the kinetic factor $\beta$, 
$\beta\approx\frac{\Delta w}{\ell}\lesssim 0.01$, 
with the aspect ratio squared, $\alpha^2\approx (\frac{\lambda_x}{\lambda_y})^2$; and (ii) the 
relative strength of step interactions,
$\frac{g_3}{g_1}$, with $(\frac{\alpha}{\nu})^2$.
In~\cite{keefeetal94} $\alpha\approx 10^{-3}$ and thus $\beta\gg \alpha^2$. 
Also, $\nu\approx 1/30$ and $\frac{g_3}{g_1}>1$~\cite{poonzandvliet}, and thus $\frac{g_3}{g_1}\gg (\frac{\alpha}{\nu})^2$.
Equation (\ref{eq:A-III}) follows, in agreement with the decay 
in~\cite{keefeetal94}. In~\cite{erlebacheretal00} $\alpha \gtrsim 10^{-1}$, 
$\nu\approx 1/15$ and $\frac{g_3}{g_1}\approx 100$~\cite{poonzandvliet}. So,
$\beta< \alpha^2$ and $\frac{g_3}{g_1}> (\frac{\alpha}{\nu})^2$. Equation (\ref{eq:A-I}) follows, 
in agreement with the inverse linear decay in~\cite{erlebacheretal00}. \looseness=-1

We now discuss observations of Ag(110)~\cite{pedemonteetal03} 
where step interactions are mainly entropic~\cite{jeongwilliams99,paietal94}. 
By $\beta< 10^{-3}$~\cite{pedemonteetal03}
and $\alpha\approx 1/15$~\cite{deMongeot}, we have $\beta< \alpha^2$.
We estimate $\frac{g_3}{g_1}$ by \mbox{$g_1=\frac{\epsilon_{\rm k}}{a a_0}-\frac{k_{\rm B}T}{a a_0}
\ln(\coth\frac{\epsilon_{\rm k}}{2k_{\rm B}T})$} and 
$g_3\approx \frac{\pi^2 a_0 k_{\rm B}T}{2 a^3} [\sinh(\frac{\epsilon_{\rm k}}{2k_{\rm B}T})]^{-2}$~\cite{jeongwilliams99},
where $\epsilon_{\rm k}$ is the kink formation energy, 
$0.04$ eV$\lesssim\epsilon_{\rm k}\lesssim 0.1$ eV~\cite{jeongwilliams99,vitosetal99},
$a=1.4$ \AA, $a_0\approx 4$ \AA, and $T=210$ K; thus, $\frac{2}{17}\lesssim\frac{g_3}{g_1}\lesssim 1$. 
With $\nu\approx 2/25$~\cite{pedemonteetal03,valbusaetal02}, $\frac{g_3}{g_1}=O(\frac{\alpha^2}{\nu^2})$; thus,
our criterion for step energetics appears inconclusive for scaling. Possible
reasons are deviations of initial profiles from sinusoidal ones and anisotropies in Ag(110), 
for which the model in~\cite{dankeretal04} may be relevant. Although further 
study of the dynamics with reliable boundary conditions
at facets is suggested, we view the condition $\beta< \alpha^2$ as an indicator of evolution
toward inverse linear decay~\cite{pedemonteetal03}. 
\looseness=-1

Our work forms a basis for a general approach to 
morphological evolution below roughening.
Extensions in 2D include the ES barrier, 
long-range step interactions, step edge diffusion, anisotropy
of step stiffness, and material deposition. 
Inclusion of the ES barrier~\cite{ES-barrier} with rates $k_{\rm u}$ and $k_{\rm d}$ 
amounts effectively to $k=2(1/k_{\rm u}+1/k_{\rm d})^{-1}$ in 
Eq.~(\ref{eq:full-PDE})~\cite{margetiskohn}.
Step-edge diffusion contributes to longitudinal fluxes but may not be important for 
Si(001), where ADL kinetics can dominate~\cite{jeongwilliams99}. Anisotropic terrace diffusion,
which is present in Si(001) and Ag(110), is not expected to alter the main decay laws presented here.

Connections of initial conditions and solutions for Eq.~(\ref{eq:full-PDE}) to actual experimental
situations have yet to be explored. Our scaling Ansatz should be tested for realistic initial
profiles. Despite mode coupling~\cite{chanetal04} caused by the nonlinear PDE (\ref{eq:full-PDE}), 
our scaling should be valid for a range of prevailing 
wavelengths~\cite{erlebacheretal00,pedemonteetal03,chanetal04}. \looseness=-1

Other predictions of our approach include  
crossovers from exponential to inverse linear profile decay via aspect-ratio changes
of the surface shape. Our work should stimulate further studies and relaxation experiments 
on surfaces below rougnening.\looseness=-1

\acknowledgments{This work
has been supported by NSF-MRSEC DMR0520471 at the University of Maryland; also, by
the U.S. Department of Energy through DE-FG02-01ER45947 via M.~J. Aziz, and by the
Harvard NSEC via H.~A. Stone.}

\end{document}